\documentclass[preprint,12pt,nopreprintline]{elsarticle}
\usepackage[utf8]{inputenc}
\usepackage[T1]{fontenc}

\newcommand{\Skip}[1]{}

\newcommand{\best}[1]{\textbf{#1}}
\newcommand{\secbest}[1]{\underline{#1}}
\newenvironment{NYC}{\color{teal}}{}

\usepackage{multirow}
\usepackage{makecell}
\usepackage{xcolor}
\usepackage{amsmath}
\usepackage{amssymb}
\usepackage{adjustbox}
\usepackage{booktabs}
\usepackage{graphicx}
\usepackage{float}

\journal{Optics and Laser Technology}

\begin{document}

\begin{frontmatter}
    
\title{CV-HoloSR: Hologram to hologram super-resolution through volume-upsampling three-dimensional scenes}

\author[1]{Youchan No}
\author[1]{Jaehong Lee}
\author[1]{Daejun Choi}
\author[2]{Dae Youl Park}
\author[1]{Duksu Kim\corref{cor1}}
\cortext[cor1]{Corresponding author}
\ead{bluekdct@gmail.com}

\affiliation[1]{organization={Korea University of Technology \& Education (KOREATECH)}, 
            city={Cheonan},
            country={Republic of Korea}}

\affiliation[2]{organization={Electronics and Telecommunications Research Institute (ETRI)}, 
            city={Daejeon},
            country={Republic of Korea}}



\begin{abstract}
    Existing hologram super-resolution (HSR) methods primarily focus on angle-of-view expansion.
    Adapting them for volumetric spatial up-sampling introduces severe quadratic depth distortion, degrading 3D focal accuracy.
    We propose CV-HoloSR, a complex-valued HSR framework specifically designed to preserve physically consistent linear depth scaling during volume up-sampling. 
    Built upon a Complex-Valued Residual Dense Network (CV-RDN) and optimized with a novel depth-aware perceptual reconstruction loss, our model effectively suppresses over-smoothing to recover sharp, high-frequency interference patterns.
    To support this, we introduce a comprehensive large-depth-range dataset with resolutions up to 4K.
    Furthermore, to overcome the inherent depth bias of pre-trained encoders when scaling to massive target volumes, we integrate a parameter-efficient fine-tuning strategy utilizing complex-valued Low-Rank Adaptation (LoRA).
    Extensive numerical and physical optical experiments demonstrate our method’s superiority.
    CV-HoloSR achieves a 32\% improvement in perceptual realism (LPIPS of 0.2001) over state-of-the-art baselines.
    Additionally, our tailored LoRA strategy requires merely 200 samples, reducing training time by over 75\% (from 22.5 to 5.2 hours) while successfully adapting the pre-trained backbone to unseen depth ranges and novel display configurations.
\end{abstract}


\end{frontmatter}

\section{Introduction}
\label{sec:intro}

Holography is a prominent field in optics, capable of representing three-dimensional (3D) scenes by exploiting light diffraction to record and reconstruct objects using holograms.  
A distinct advantage of holography is its ability to display full 3D scenes directly through holographic content, without the need for additional viewing devices such as head-mounted displays (HMDs).  
However, recording real-world scenes using holograms requires precise manipulation of optical devices and remains inherently challenging, as the medium is typically not reusable for capturing new scenes.

Computer-generated holography (CGH)~\cite{slinger2005computer} overcomes these physical constraints by generating holograms through computer simulations.
It enables efficient hologram synthesis from digital graphical representations—such as point clouds and meshes—that can be easily stored and processed.
Typical CGH methods include layer-based~\cite{zhao2015accurate,kim2017ultrafast}, point-based~\cite{su2016fast,maimone2017holographic,symeonidou2018colour}, mesh-based~\cite{askari2017occlusion,ko2017speckle,yeom2022efficient}, and light field–based approaches~\cite{park2019non,park2020hologram}.
However, CGH encounters substantial computational challenges due to discrete sampling, resulting in a trade-off between reconstruction quality and computational cost.
Meeting the Nyquist sampling criterion for high-frequency components demands extensive computational resources.
Although several approaches, such as look-up table (LUT) methods~\cite{Lookup1, Lookup2, Lookup3}, have been proposed to enhance efficiency, a fundamental limitation remains: computational complexity scales quadratically with hologram resolution.

To overcome these computational bottlenecks~\cite{shimobaba2018efficient,shimobaba2019computer,blinder2019efficient,lee2021out,lee2023out,lee2024combo}, recent research has increasingly adopted data-driven approaches that leverage deep learning for efficient, high-quality hologram generation~\cite{peng2020neural,ishii2023multi}. For instance, Tensor Holography~\cite{shi2021towards} employs a convolutional neural network (CNN) to efficiently generate full high-definition (FHD) holograms in real time.
The accompanying MIT-CGH-4K dataset has significantly accelerated research progress in this domain~\cite{mithologen1, mithologen2, mithologen3, mithologen4}, including the rapidly emerging field of hologram super-resolution (HSR)~\cite{jee2022hologram,lee2024holosr,no2024h2hsr}.


High-resolution holograms are essential because their spatial resolution dictates both the physical display size and the angle of view (AoV).
However, unlike standard 2D images, simple resolution enhancement techniques (e.g., bicubic interpolation) cannot be directly applied to complex-valued holograms.
Naive spatial scaling alters the underlying fringe frequencies, introducing severe physical artifacts—most notably depth distortion, where the reconstructed 3D volume expands quadratically rather than linearly with the scaling factor~\cite{park2020generation}.
While Park et al.~\cite{park2020generation} addressed this distortion by interpolating in the light-field domain, the process is computationally heavy, requiring multiple transformations between hologram and light-field representations. 


These challenges motivate the development of direct deep learning-based HSR techniques.
HSR typically falls into two distinct categories: Angle-of-View (AoV) up-scaling and volume up-sampling.
AoV up-scaling involves increasing the pixel count while decreasing the pixel pitch, resulting in a wider viewing angle while maintaining the original reconstructed image size.
Conversely, volume up-sampling increases the spatial resolution while keeping the pixel pitch fixed.
This operates analogously to 2D image super-resolution, aiming to linearly expand the reconstructed scene's physical volume or display size without suffering from the aforementioned quadratic depth distortions.


HSR has been extensively investigated in microscopy, utilizing both optical/numerical techniques~\cite{micro_optical1, micro_optical2, micro_optical3, micro_numerical1, micro_numerical2, micro_numerical3} and deep learning–based methods~\cite{micro_dnn1, micro_dnn2, micro_dnn3}. 
In this context, coherent imaging systems are widely used, and the primary goal is often to recover missing information (e.g., phase) resulting from sensor chip limitations, rather than focusing on the full-color holograms typically generated through CGH.

Recently, several studies have introduced HSR methods for full-color holograms using deep learning techniques.
The MIT-CGH-4K dataset has played a pivotal role in these advancements by providing high-quality CGH data at two resolutions: a low-resolution (LR) hologram of $192 \times 192$ pixels with a pixel pitch of 16 µm, and a high-resolution (HR) hologram of $384 \times 384$ pixels with a pixel pitch of 8 µm.
Lee et al.~\cite{lee2024holosr} proposed a deep learning–based HSR network that takes an RGB-D image as input and generates an up-scaled hologram as output.
Jee et al.~\cite{jee2022hologram} developed a Dual-GAN–based HSR framework that reconstructs both amplitude and phase components, producing high-quality super-resolved holograms.
Furthermore, a hologram-to-hologram SR method (H2HSR)~\cite{no2024h2hsr} was proposed, which adapts an image super-resolution network to the HSR task while employing the loss functions introduced in Tensor Holography~\cite{shi2021towards}.
H2HSR achieved up to 8.46 dB in PSNR and 9.30 in SSIM, significantly outperforming conventional interpolation-based methods in terms of reconstruction quality.
While previous HSR studies have primarily focused on angle-of-view (AoV) up-scaling to widen viewing angles, the problem of volume up-sampling—which aims to preserve linear depth scaling in reconstructed 3D scenes—remains largely unexplored.
Moreover, most existing works have been limited to relatively small resolutions (e.g., $192^2$ to $384^2$) and shallow depth ranges (e.g., $-3~mm$ - $3~mm$), constrained by the characteristics of the MIT-CGH-4K dataset.




To bridge this gap, we propose a novel deep learning framework specifically designed for hologram volume up-sampling.
Our network, built upon a Complex-Valued Residual Dense Network (CV-RDN), processes holograms directly in the complex domain to preserve physical wavefield interactions (Sec.~\ref{sec:method-network}).
To overcome the limitations of existing datasets, we construct and release a comprehensive, large-depth-range hologram dataset containing 4,000 paired samples with resolutions up to $4096^2$ (Sec.~\ref{sec:dataset}).
We further introduce a cropping-based training strategy (Sec.~\ref{subsec:cropping}) paired with a depth-aware perceptual loss (Sec.~\ref{subsec:loss}) to efficiently optimize the network for physically consistent 3D reconstructions on limited GPU memory. 
Finally, to address the inherent depth bias of pre-trained encoders when applied to massive target resolutions, we introduce a parameter-efficient fine-tuning strategy utilizing complex-valued Low-Rank Adaptation (LoRA), ensuring rapid generalization to unseen depth ranges (Sec.~\ref{sec:LoRA}).

Extensive numerical simulations and physical optical experiments demonstrate that our method successfully generates volumetric super-resolved holograms free of depth distortion (Sec.~\ref{sec:experiments}).
Compared to state-of-the-art spatial up-sampling baselines, our proposed CV-RDN achieves superior perceptual realism and structural fidelity (Sec.~\ref{subsec:evaluation}).
Specifically, our framework achieves a Learned Perceptual Image Patch Similarity (LPIPS) score of 0.2001 on the HologramSR dataset (a 32\% improvement over the prior SOTA baseline), producing sharp textures and natural defocus blur across expansive 3D spaces.
Furthermore, our tailored LoRA-based adaptation strategy proves highly data- and compute-efficient; by utilizing merely 200 training samples, it reduces the computational training time from 22.5 hours to just 5.2 hours (an over 75\% reduction) while preserving state-of-the-art perceptual quality (Sec.~\ref{subsec:result_LORA}).
This drastically lowers the computational burden required to adapt a pre-trained backbone to novel optical display configurations and unseen depth ranges.




\section{Dataset Generation}
\label{sec:dataset}


To train and evaluate hologram volume up-scaling, we require paired holograms spanning diverse resolutions and depth variations while maintaining physically consistent depth scaling.
Although the MIT-CGH-4K dataset~\cite{shi2021towards} provides high-quality CGH data for deep learning, it is primarily tailored to AoV up-scaling: paired holograms vary in pixel pitch and are limited to small resolutions (e.g., $192^2$ and $384^2$) with a short depth range (e.g., $-3~mm$ - $3~mm$).
Such characteristics are not well suited for volume up-scaling, where the pixel pitch is kept fixed and the reconstructed scene volume should expand while preserving linear depth scaling.


\begin{figure*}[t!]
    \centering
    \includegraphics[width=\linewidth]{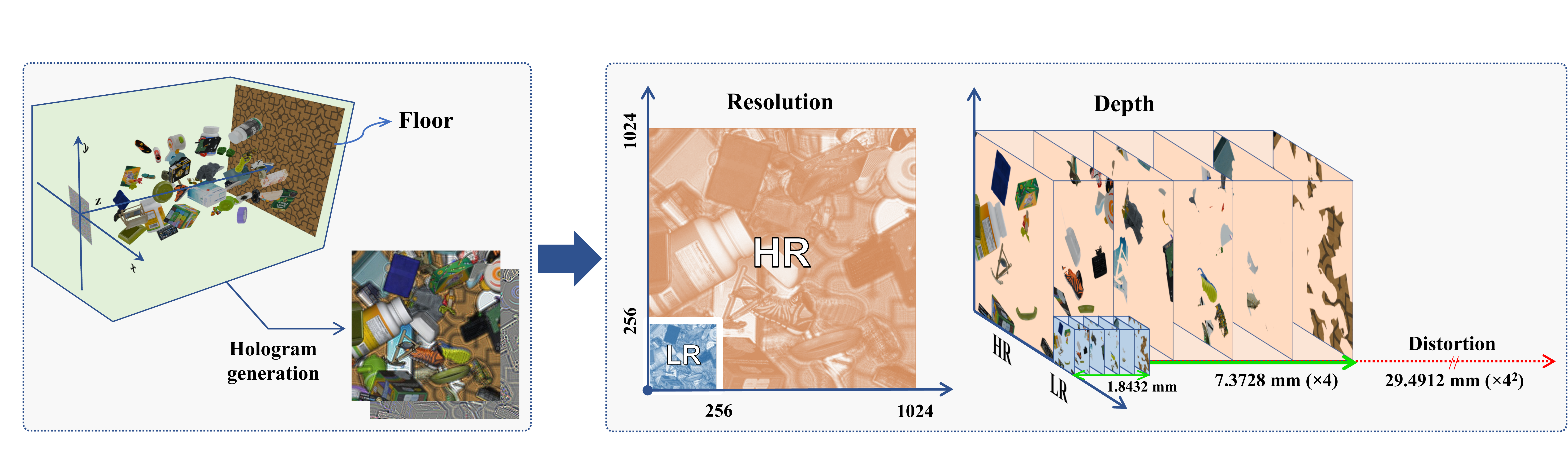}
    \caption{The overview of hologram SR dataset with a pair of LR and HR.
    }
    \label{fig:dataset}
\end{figure*}

To address this gap, we generate a new dataset specifically designed for volume up-scaling (Fig.~\ref{fig:dataset}).
Using an advanced silhouette-masking layer-based CGH (AP-LBM)~\cite{koreatech-cgh}, we synthesize paired holograms under a fixed pixel pitch and an extended depth configuration, enabling volumetric training and evaluation across a wide range of hologram resolutions.

\paragraph{Depth-range setup}
Selecting an appropriate depth range is crucial for stable training and generalization.
If the propagation distance is excessively large, diffraction patterns become overly sharp, and convolution-based networks with limited receptive fields tend to yield over-smoothed reconstructions.
Conversely, if the depth range is too short, the model observes only a narrow distribution of phase patterns, limiting its ability to handle unseen depths.

A key difficulty is that hologram phase is typically wrapped within $[-\pi, \pi]$ regardless of the scene depth, which weakens depth discriminability from the input alone.
Moreover, even when training data span a wide depth range, receptive-field limitations in deep networks can still hinder accurate reconstruction of distant content~\cite{yu2023deep,zhong2023real,jin2024vision,liu2025propagation}.

\begin{table*}[th]
    \footnotesize 
    \centering 
    \setlength{\tabcolsep}{2.5pt}
    \renewcommand{\arraystretch}{1.1}
    \begin{tabular}{c|c|r|r|r|r|r|r|r}
        \Xhline{1.0pt}
        \multicolumn{2}{c|}{\multirow{2}{*}{\textbf{Distance (mm)}}} & \multicolumn{7}{c}{\textbf{Resolution ($M$)}} \\ 
        \cline{3-9}
        \multicolumn{2}{c|}{} & $256^2$ & $384^2$ & $512^2$ & $1024^2$ & $1536^2$ & $2048^2$ & $4096^2$ \\
        \hline
        \multirow{3}{*}{Maximum} 
            & \textcolor{red}{$R$}             & \textbf{10.3596} & \textbf{15.5394} & \textbf{20.7192} & \textbf{41.4384} & \textbf{62.1576} & \textbf{82.8768} & \textbf{165.7534} \\
            & \textcolor{green!60!black}{$G$}  & 12.4386 & 18.6579 & 24.8774 & 49.7548 & 74.6316 &  99.5094 & 199.0190 \\
            & \textcolor{blue}{$B$}            & 14.7168 & 22.0752 & 29.4336 & 58.8670 & 88.3008 & 117.7342 & 235.4684 \\
        \hline
        \multirow{1}{*}{Dataset} 
            & $RGB$ & \textbf{1.8432} & \textbf{2.7648} & \textbf{3.6864} & \textbf{7.3728} & \textbf{11.0592} & \textbf{14.7456} & \textbf{29.4912} \\
        \Xhline{1.0pt}
    \end{tabular}
    \caption{Maximum propagation distances under the angular spectrum method (ASM) for different hologram resolutions and wavelengths (638~nm red, 532~nm green, and 450~nm blue). Values labeled as \textit{Dataset} denote the maximum distances adopted for hologram dataset generation.
    }
    \label{tab:max_distance_asm}
\end{table*}

Previous studies~\cite{matsushima2020introduction,he2020optimal} often determine depth ranges from hologram resolution and pixel pitch, and Table~\ref{tab:max_distance_asm} summarizes the corresponding theoretical maximum propagation distances under the angular spectrum method (ASM).
However, adopting these theoretical maxima can exacerbate over-smoothing and reduce generalization.
Therefore, we use a more practical depth configuration based on the physical aspect ratio of the hologram volume, adopting a commonly used ratio of $1:1:2$ ($x$:$y$:$z$), where the depth is set to twice the lateral size.

\paragraph{Zero-point hologram}
A common strategy to alleviate artifacts associated with depth ranges is to place the hologram plane at the mid-point of the 3D scene~\cite{shi2021towards}, which reduces sharp diffraction patterns from far-depth layers and mitigates over-smoothing.
However, mid-point placement requires prior knowledge of the scene depth range; when the input hologram’s depth extent is unknown, the appropriate mid-point propagation is difficult to determine.

To avoid this dependency, we instead place the hologram plane at the zero position along the depth axis (0~mm), which we refer to as a \emph{zero-point hologram}.
Training on zero-point holograms allows the network to process scenes spanning from the zero plane to the maximum supported depth at a given resolution, without requiring explicit depth-range information at inference time.

\paragraph{Dataset specification}
Our dataset contains 4,000 unique 3D scenes and 4,000 paired samples for volume up-scaling.
We consider resolutions ${256^2, 512^2, \dots, 4096^2}$ with a fixed pixel pitch of 3.6~$u$m.
Each sample includes complex-valued RGB holograms (amplitude and phase) generated under the ASM, along with the corresponding depth map.
The scene depth spans 1.84--29.49~mm and is discretized into 4,096 depth planes.
We split the dataset into train/validation/test sets using a ratio of 3,800:100:100, ensuring no scene overlap across splits.

\section{Complex-valued Hologram Super-resolution Network}
\label{sec:method-network}

In this section, we present a complex-valued hologram super-resolution network for volumetric up-scaling.
Our model operates directly on complex RGB holograms and employs a complex-valued residual dense backbone with a sub-pixel upsampling head.
To enable training on high-resolution holograms under limited GPU memory, we adopt a patch-based training strategy and optimize the network using a physically informed loss design.

\subsection{Network Overview}

\begin{figure*}
    \centering
    \includegraphics[width=0.95\linewidth]{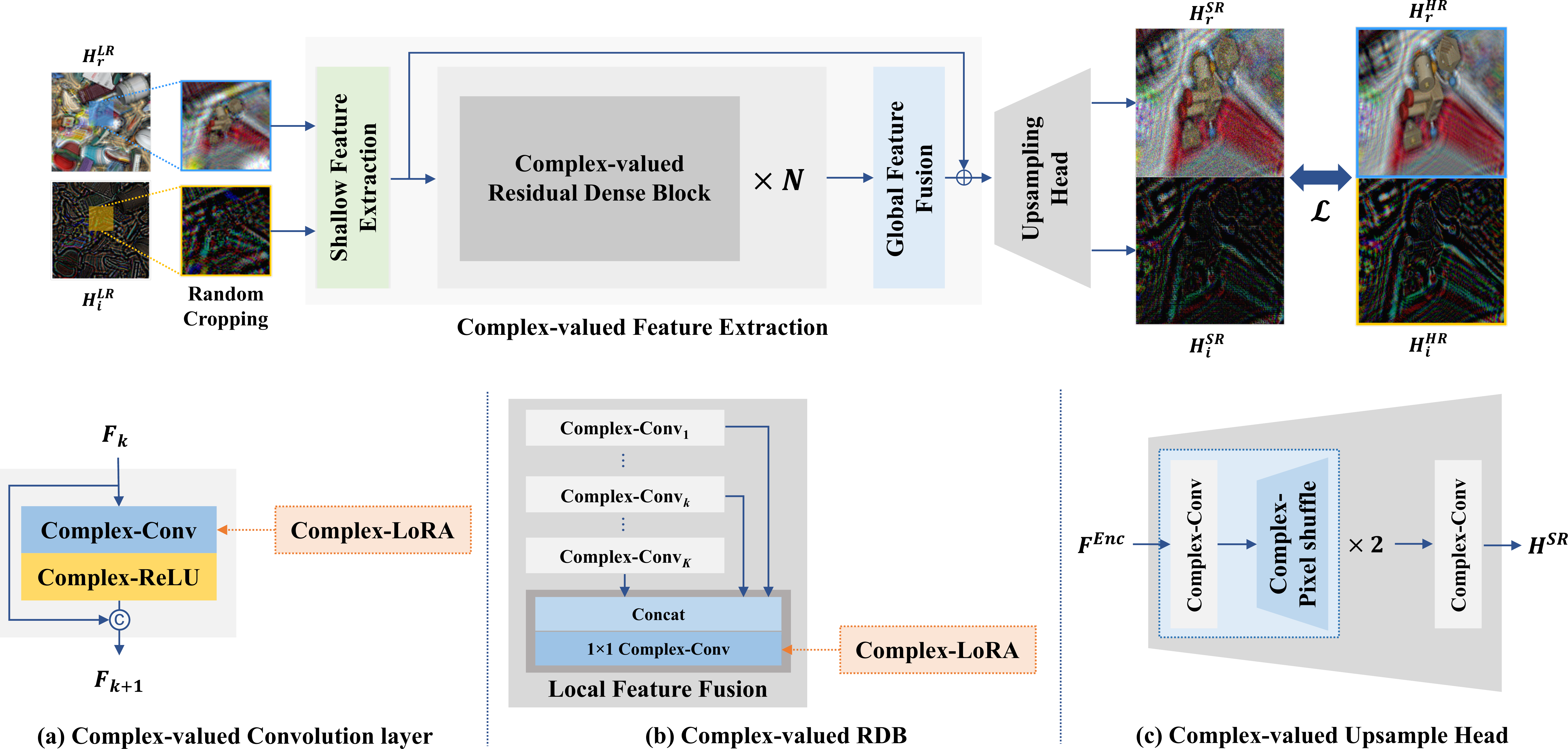}
    \caption{The overview of proposed network for hologram super-resolution.
    } 
    \label{fig:overview-proposed-method}
\end{figure*}

We address hologram volume up-scaling by learning a direct mapping from a low-resolution (LR) complex hologram to its high-resolution (HR) counterpart, while preserving reconstruction fidelity and depth-consistent scaling in the reconstructed 3D scene.
In our setting, both input and output are complex-valued RGB holograms, where each channel encodes a wavelength-dependent complex wavefront (Sec.~\ref{subsec:complex}).
Our objective is to super-resolve the hologram itself so that subsequent numerical propagation yields volumetrically consistent reconstructions without introducing depth distortions.

Fig.~\ref{fig:overview-proposed-method} illustrates the overall pipeline.
During training, we apply random cropping (Sec.~\ref{subsec:cropping}) to reduce memory usage and enable learning from high-resolution holograms; this is important because hologram resolution directly affects the depth range that the network can cover.
The cropped LR hologram is first processed by a shallow feature extraction module composed of complex-valued convolution layers, producing low-level complex feature maps.
These features are then passed through a stack of complex-valued residual dense blocks (CV-RDBs), which progressively refine representations through dense connections and residual learning (Sec.~\ref{subsec:CV-RDN}).
The resulting multi-level features are aggregated via a global feature fusion module to form a compact, high-capacity representation for up-scaling.

To generate the HR hologram, we employ a complex sub-pixel upsampling head that increases spatial resolution using complex-valued convolution followed by pixel shuffle.
The network output is supervised against the corresponding cropped HR hologram using a composite objective function $\mathcal{L}$, which enforces both complex-domain fidelity and reconstruction-aware consistency (Sec.~\ref{subsec:loss}).
We adopt a regression-oriented formulation rather than a generative approach, as hologram super-resolution requires strict fidelity to the input wavefront to ensure physically plausible propagation and stable 3D reconstruction across depth.

\subsection{Complex-Valued Representation and Operations}
\label{subsec:complex}

Complex-valued hologram super-resolution benefits from modeling holograms in their native complex domain, since both amplitude and phase jointly determine wave propagation and reconstruction.
In this work, we represent a hologram as a complex field $H = H_r + iH_i$,
where $H_r$ and $H_i$ denote the real and imaginary components, respectively.
Although holograms are often visualized as amplitude and phase via Euler’s rule, we operate on the complex representation to avoid ambiguity introduced by phase wrapping and to preserve a physically meaningful form for learning.
For full-color holograms, we treat the input as complex-valued RGB channels, i.e., three wavelength-dependent complex fields that are processed jointly by the network.

To learn directly from complex inputs, we employ complex-valued convolution layers.
Let $X = X_r + iX_i$ be a complex feature map and $W = W_r + iW_i$ a complex convolution kernel. The complex convolution output $Y = X * W$ is computed as
\begin{equation}
Y_r = X_r * W_r - X_i * W_i,\qquad
Y_i = X_r * W_i + X_i * W_r,
\label{eq:complex_conv}
\end{equation}
which corresponds to the standard complex multiplication rule and is illustrated in Fig.~\ref{fig:overview-proposed-method}.
In practice, this operation is implemented using four real-valued convolutions, since each complex convolution decomposes into coupled real/imaginary branches.
Compared to real-valued networks that concatenate amplitude and phase along the channel dimension, complex-valued convolutions explicitly model interactions between real and imaginary components through learned kernels, providing a more direct mechanism for capturing phase-sensitive features relevant to hologram reconstruction.

After each complex convolution, we apply a component-wise nonlinearity.
Specifically, we use the standard ReLU independently on the real and imaginary parts (i.e., $\mathrm{ReLU}(X_r) + i\,\mathrm{ReLU}(X_i)$).
While alternative complex activations such as modReLU~\cite{arjovsky2016unitary} and zReLU~\cite{guberman2016complex} have been proposed, we found component-wise ReLU to provide stable and effective optimization in our setting.
Residual connections are applied in the complex domain in the same manner as in real-valued networks, by element-wise addition of complex feature maps.

Overall, these representation and operator choices allow the network to process holograms in a physically consistent complex form, while keeping the subsequent architectural description clean and modular.

\subsection{CV-RDN Architecture and Upsampling Module}
\label{subsec:CV-RDN}

Our CV-RDN architecture follows the overall design of residual dense networks (RDNs)~\cite{zhang2018residual,lim2017enhanced} while operating entirely in the complex domain.
As shown in Fig.~\ref{fig:overview-proposed-method}, the network consists of shallow feature extraction, a stack of complex-valued residual dense blocks (CV-RDBs), global feature fusion, and a complex sub-pixel upsampling head that produces the super-resolved hologram.

\paragraph{Shallow feature extraction}
Given an input complex hologram $H^{LR}$, we first extract low-level features using shallow complex-valued convolution layers (e.g., 2~layers), producing an initial feature map $F_0$.
This stage encodes basic fringe patterns and local phase variations into a compact complex feature representation for subsequent refinement.

\paragraph{Complex-valued residual dense blocks}
We adopt residual dense blocks as the main backbone module.
Each CV-RDB contains multiple complex-valued convolution layers connected through dense skip connections, enabling feature reuse and strengthening gradient flow.
Within a block, outputs from intermediate layers are concatenated and compressed using a local feature fusion (LFF) layer.
In addition, residual learning is applied at the block level by adding the block input to the fused output, which stabilizes training and promotes progressive refinement of complex-domain features.

\paragraph{Global feature fusion}
Let $F_1,\ldots,F_B$ denote the outputs of the $B$ CV-RDBs.
Following the RDN design, we concatenate these block outputs and apply a global feature fusion module composed of complex-valued convolution layers.
This module integrates multi-level representations and produces a globally fused feature map.
A global residual connection is further employed by adding the shallow feature $F_0$ to the fused features, encouraging the network to learn high-frequency residual components required for hologram super-resolution.

\paragraph{Upsampling head with complex sub-pixel convolution}
To generate a high-resolution hologram, we use a sub-pixel upsampling head based on pixel shuffle~\cite{subpixel}.
Specifically, a complex-valued convolution expands the channel dimension from $C$ to $s^2C$, where $s$ denotes the up-scaling factor.
The pixel shuffle operation then rearranges these expanded channels into a higher spatial resolution feature map.
For complex-valued features, the same rearrangement is applied consistently to the real and imaginary components.
This design increases spatial resolution while keeping the network fully convolutional and efficient.

\paragraph{Output layer}
Finally, a complex-valued convolution layer maps the upsampled features to the output hologram $H^{SR}$ in complex form (real and imaginary components).
This output is used as the super-resolved hologram for subsequent numerical propagation and reconstruction.

\subsection{Handling Cropping Artifacts in Hologram Super-resolution}
\label{subsec:cropping}


Patch-based training, or cropping, is a widely adopted strategy in deep learning-based holography---such as in compression, microscopy, and phase retrieval~\cite{holocrop1, holocrop2, holocrop3, holocrop4}---to reduce memory consumption, augment data, and improve generalization.
In our hologram super-resolution framework, cropping is similarly essential for processing high-resolution holograms while balancing depth coverage and GPU memory limits (Sec.~\ref{sec:dataset}).
However, when applied to tasks requiring numerical wave propagation, cropping introduces specific physical challenges.
Because the angular spectrum method (ASM) implicitly assumes periodic boundary conditions, restricting a hologram's spatial support via cropping emphasizes boundary discontinuities.
This leads to severe ringing artifacts in the reconstructed field that worsen at larger propagation distances.

\begin{figure*}
\centering
\includegraphics[width=0.95\linewidth]{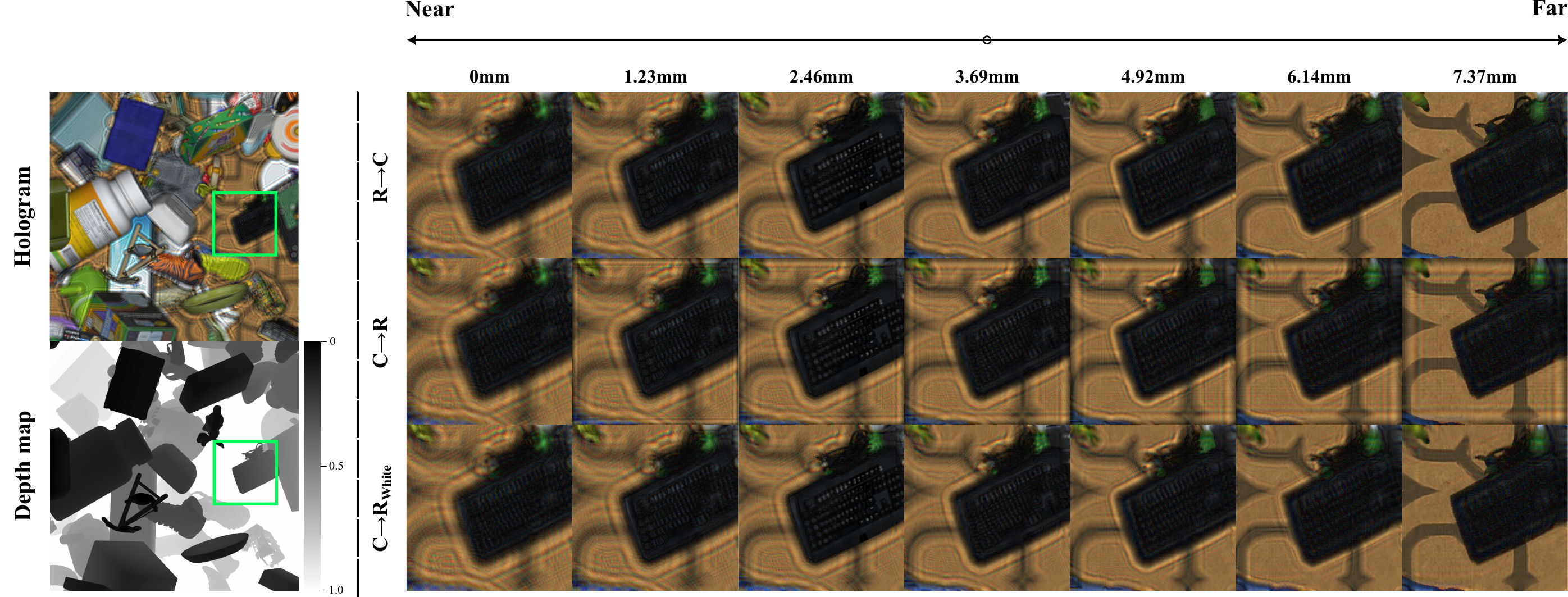}
\caption{Cropping-induced ringing artifacts in ASM reconstruction and the effect of the white-hologram formulation.}
\label{fig:cropping-comparison}
\end{figure*}

Conventional approaches mitigate these boundary effects using apodization windows (e.g., Tukey or cosine tapers)~\cite{apodization_tukey, apodization_cosine} or a white-hologram formulation~\cite{apodization_white}.
As illustrated in Fig.~\ref{fig:cropping-comparison}, replacing the cropped hologram's amplitude with a constant (white) amplitude while preserving its phase (C$\rightarrow$R$_{\text{white}}$) visually reduces boundary ringing at intermediate distances compared to the direct propagation of a cropped patch (C$\rightarrow$R).

However, a critical observation in our hologram-to-hologram framework is that explicitly mitigating these artifacts is largely unnecessary for loss optimization.
Because the SR and HR holograms are propagated using identical ASM parameters, the boundary-induced ringing artifacts appear symmetrically in both reconstructions.
Consequently, these artifacts effectively cancel out during the SR--HR loss comparison.
This contrasts sharply with self-supervised 2D hologram frameworks~\cite{peng2020neural, chakravarthula2019wirtinger, zhang20173d}, where the reconstructed field is compared against an artifact-free RGB image captured at a single depth plane, making boundary mitigation techniques strictly necessary.

Finally, this cropping behavior directly informs our configuration of reconstruction planes for ASM-based supervision.
Because a cropped patch only contains partial spatial information, uniformly sampling planes at very far propagation distances over-emphasizes out-of-focus regions that provide limited perceptual feedback for training.
Therefore, when computing reconstruction-based losses, we specifically sample planes around the in-focus depth interval that is reliably represented by the cropped hologram, as detailed in Sec.~\ref{subsec:loss}.

\subsection{Loss Function}
\label{subsec:loss}


To train the proposed network, we employ a composite objective function that balances numerical signal accuracy with perceptual reconstruction quality:
$\mathcal{L}_{\mathrm{total}} = \mathcal{L}_{\mathrm{data}} + \lambda \mathcal{L}_{\mathrm{ASM\text{-}LPIPS}}$, where $\lambda$ is a balancing coefficient.

\paragraph{Data fidelity loss}
The data fidelity term enforces strict numerical regression between the predicted hologram $\hat{y}$ and the ground-truth $y$ in the complex domain.
This direct supervision is critical because hologram volume up-sampling requires a precise linear depth scaling--mapping the spatial expansion directly to the depth axis--rather than the quadratic expansion inherent to standard physical scaling.
We apply the $L_1$ loss independently to the real and imaginary components:
\begin{equation}
    \mathcal{L}_{\mathrm{data}} = \frac{1}{2} \left( \mathcal{L}_{1}(y_r , \hat{y}_r) + \mathcal{L}_{1}(y_i , \hat{y}_i) \right).
    \label{eq:data-fidelity-loss}
\end{equation}

\paragraph{Depth-aware perceptual reconstruction loss}
While $\mathcal{L}_{\mathrm{data}}$ ensures signal consistency, complex-domain pixel-wise losses often favor "averaged" solutions for ill-posed problems, leading to over-smoothed holograms that lack high-frequency interference details.
To guarantee physically consistent and perceptually faithful 3D reconstructions, we incorporate a reconstruction-based perceptual loss.
Using the ASM~\cite{matsushima2009band}, we propagate the holograms to a set of numerical reconstruction planes and evaluate them using Learned Perceptual Image Patch Similarity (LPIPS)~\cite{LPIPS}.

As discussed in Sec.~\ref{subsec:cropping}, evaluating reconstructions at arbitrary distances is uninformative for cropped patches.
Therefore, we restrict the evaluation to a valid depth interval $[z_{\min}, z_{\max}]$ present within the specific patch.
We uniformly partition this interval into $N$ sub-intervals and sample one propagation distance $z_i$ from each:
\begin{equation}
    z_i \sim \mathcal{U}\!\left(z_{\min} + (i-1)\Delta z,\; z_{\min} + i\Delta z\right), \quad \Delta z = \frac{z_{\max}-z_{\min}}{N}, \quad i=1,\ldots,N.
    \label{eq:z-stratified-sampling}
\end{equation}

Let $r_{z_i} = \mathbf{ASM}(y, z_i)$ and $\hat{r}_{z_i} = \mathbf{ASM}(\hat{y}, z_i)$ denote the reconstructed fields at depth $z_i$.
The ASM-LPIPS loss is defined as the average perceptual distance across these planes:
\begin{equation}
    \mathcal{L}_{\mathrm{ASM\text{-}LPIPS}} = \frac{1}{N} \sum_{i=1}^{N} \mathrm{LPIPS}(r_{z_i}, \hat{r}_{z_i}).
    \label{eq:asm-lpips-loss}
\end{equation}

Unlike Tensor Holography~\cite{shi2021towards}, which employs pixel-wise depth weighting to emphasize focused regions, our approach utilizes uniform slicing across the depth interval. Since LPIPS evaluates similarity in a deep feature space rather than at the pixel level, a strict depth-weighting strategy is less compatible. However, we observe that uniform sampling effectively provides distributed supervision across all hologram pixels, ensuring that both in-focus details and out-of-focus blur characteristics are faithfully preserved throughout the entire volume.

\Skip{
\begin{NYC}
Let $r_{z_i} = \mathbf{ASM}(y, z_i)$ and $\hat{r}_{z_i} = \mathbf{ASM}(\hat{y}, z_i)$ denote the reconstructed fields at the depth plane $z_i$.
Although scene geometry is not uniformly distributed along the depth axis, uniformly sliced depth planes still provide supervision to all hologram pixels through the loss function.
Tensor Holography~\cite{shi2021towards} employs pixel-wise depth weighting to emphasize focused pixels. However, such a strategy is not directly compatible with LPIPS, which evaluates perceptual similarity in a deep feature space rather than at the pixel level.
Nevertheless, we observe that the use of uniformly sliced depth planes improves the visual quality of reconstructed fields.

To this end, we utilize a precomputed depth range for ASM-based reconstruction.
The ASM-LPIPS loss is defined as
\begin{equation}
    \mathcal{L}_{\mathrm{ASM\text{-}LPIPS}} = \sum_{i=1}^{N} \mathrm{LPIPS}(r_{z_i}, \hat{r}_{z_i}),
    \label{eq:asm-lpips-loss}
\end{equation}
where $N$ denotes the number of uniformly sampled depth planes.
Each reconstruction is evaluated using LPIPS, and the resulting values are summed and averaged over the number of depth planes.
\end{NYC}
}

\subsection{Parameter-Efficient Depth Range Adaptation}
\label{sec:LoRA}


As deep learning-based methods generally perform well only for inputs similar to the training dataset, CGH networks exhibit a strong dependency on the dataset's specific depth configuration.
Prior deep learning-based CGH methods usually assume a narrow depth interval during training.
This is feasible for practical applications such as near-eye holography~\cite{peng2020neural, shi2021towards}, which typically relies on a fixed optical geometry, or digital holographic microscopy~\cite{micro_dnn1, micro_dnn2, micro_dnn3}, which is often conducted under controlled acquisition settings.

In contrast, we target a generalizable hologram super-resolution network capable of super-resolving holograms whose reconstructed scenes exhibit a linearly extended depth range according to the spatial up-scaling factor. 
However, we found that our base SR network struggles to handle holograms with depth intervals (i.e., fringe statistics) that differ from the training dataset.

\begin{figure*}
\centering
\includegraphics[width=0.95\linewidth]{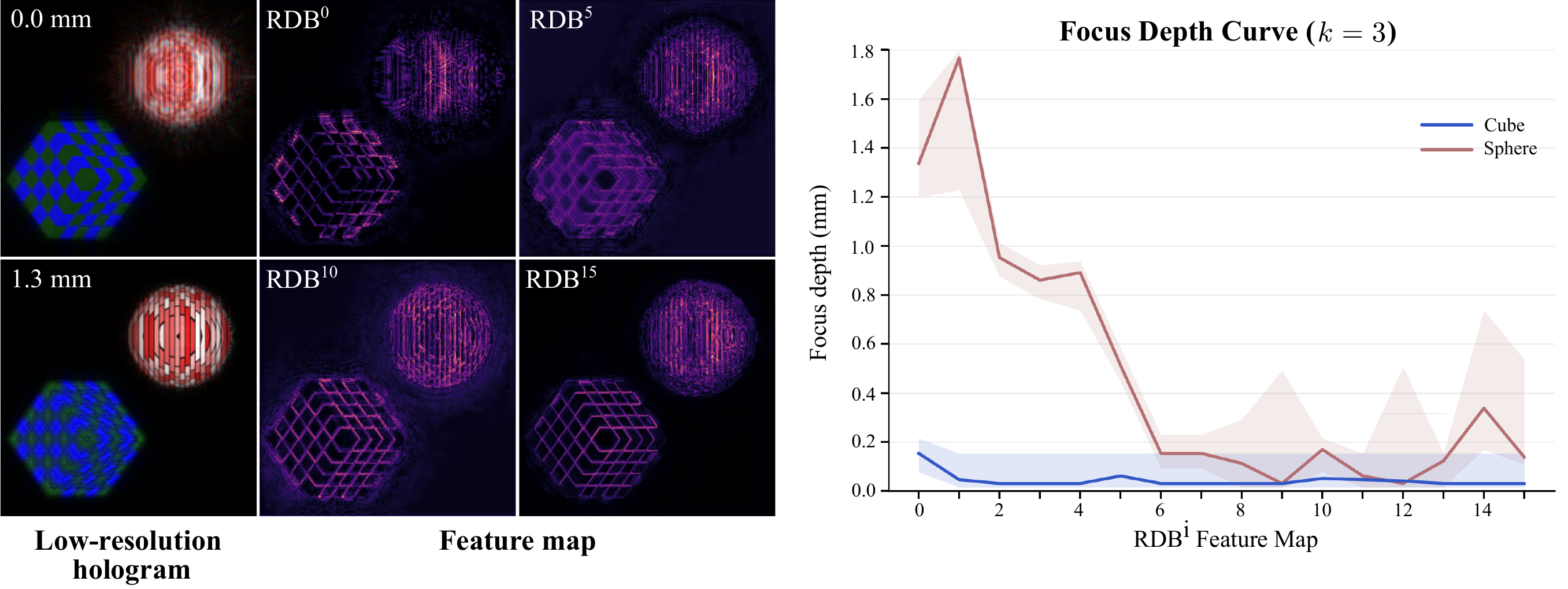}
\caption{Analysis of depth bias in the pretrained encoder.
(Left) Numerical reconstructions of the input low-resolution hologram at 0.0~mm and 1.3~mm, showing a cube and a sphere, respectively.
(Middle) Reconstructed amplitudes of intermediate complex feature maps extracted from selected residual dense blocks (RDBs).
(Right) The estimated focus depth curve demonstrates that while the cube remains in focus near 0.0~mm, the sphere's focus depth progressively shifts from 1.3~mm toward 0.0~mm across the RDBs, confirming the encoder's inherent bias toward a narrow depth interval.}
\label{fig:method-general}
\end{figure*}

To investigate this issue, we analyze how depth-related characteristics evolve across the pretrained encoder's intermediate feature maps.
As illustrated in Fig.~\ref{fig:method-general}, we process a hologram containing a cube at 0.0 mm and a sphere at 1.3 mm, estimating the focus depth for each object by reconstructing images from the feature maps at the end of various residual dense blocks (RDBs).
Specifically, we compute the Sobel gradient-magnitude edge map $E_{\mathrm{ref}}$ of the object from an RGB reference, and compare it against the reconstructed edge map $E_i(z)$ over multiple propagation distances $z \in \mathcal{Z}$ using normalized cross-correlation ($\mathrm{NCC}_i(z)$).
To reduce sensitivity to discrete sampling, the block-wise focus depth is defined as the peak location $\hat{z}_i=\arg\max_{z\in\mathcal{Z}}\mathrm{NCC}_i(z)$, averaged over the top-$k$ candidates.

The resulting focus depth curve in Fig.~\ref{fig:method-general} demonstrates that while the cube remains near 0.0 mm, the sphere's estimated focus depth shifts progressively from 1.3 mm toward 0.0 mm across the RDB feature maps. 
This confirms that the pretrained encoder inherently biases intermediate fringe representations toward the specific narrow depth interval observed during training.

To handle new depth ranges without the massive computational cost of retraining the entire network, we adopt Low-Rank Adaptation (LoRA)~\cite{LoRA} as a parameter-efficient fine-tuning strategy.
LoRA freezes the pretrained weights $W$ and learns a low-rank update:
\begin{equation}
    W' = W + \Delta W, \quad \Delta W = BA,
    \label{eq:lora}
\end{equation}
, where $A$ and $B$ are trainable low-rank matrices.
Guided by our physical analysis of the encoder's depth bias, we uniquely insert these LoRA modules into the complex-valued convolution layers inside the RDBs and the local feature fusion layer.
By fine-tuning only these complex-valued low-rank parameters, we efficiently adapt the depth-dependent mapping to new acquisition or display configurations while fully leveraging the pretrained backbone.


\section{Experiments}
\label{sec:experiments}

\subsection{Experimental Setup}
\label{subsec:experimental-setup}

\paragraph{Datasets}
Our primary synthetic dataset (i.e., HologramSR), detailed in Sec.~\ref{sec:dataset}, consists of 4,000 holograms split into 3,800 for training, 100 for validation, and 100 for testing.
The models are trained to perform $4\times$ spatial up-scaling and support multi-resolution processing, handling low-resolution (LR) inputs of $256 \times 256$, which yield corresponding high-resolution (HR) outputs of $1024 \times 1024$.

To evaluate model generalization on real-world and continuous scenes, we adapted two standard image super-resolution datasets: Big Buck Bunny (BBB)~\cite{roosendaal2008big} and RealSR~\cite{cai2019toward}.
To convert these 2D image datasets into 3D holograms, we utilized monocular depth estimation~\cite{yang2024depth, chen2025video} to extract corresponding depth maps, explicitly scaling them to match the depth range of our synthetic training set (0.0 mm to 7.3728 mm).
For the BBB dataset, we extracted 240 frames (representing 10-second video segments at 24 Hz) at 1920 × 1080 resolution, and generated the corresponding LR holograms by downscaling the spatial resolution by a factor of four (480 × 270).
For the RealSR dataset, we synthesized 100 holograms from real-world, high-resolution photographs captured with diverse aspect ratios.

\paragraph{Implementation details}
All networks were implemented in PyTorch and trained using four NVIDIA RTX A6000 (48 GB) GPUs.
During training, we extracted cropped patches of 64 × 64 from the LR inputs and 256 × 256 from the HR targets.
The models were trained for 400 epochs with a total batch size of 32.
We employed the AdamW optimizer~\cite{loshchilov2017decoupled} with momentum parameters $\beta_1=0.9$ and $\beta_2=0.99$, and a weight decay of $1.0\times10^{-5}$.
The learning rate was modulated using a CosineAnnealingLR scheduler~\cite{loshchilov2016sgdr}.
For our proposed complex-valued architecture, the initial learning rate was set to $4.0\times10^{-4}$.

\paragraph{Evaluation models}
Our proposed model is a complex-valued extension of the RDN architecture (CV-RDN).
To maintain a comparable parameter count to real-valued baselines, we halved the channel dimensions in our complex-valued version.
For ablation and upper-bound analysis, we also trained a strictly real-valued RDN under identical configurations, as well as a high-capacity variant of our proposed network (CV-RDN-H) with doubled channel dimensions.

To verify the depth distortion phenomenon discussed in Sec.~\ref{sec:intro}, we included bicubic interpolation as a fundamental baseline.
Specifically, bicubic interpolation was examined under two settings: with calibration (w/ Calib.) and without calibration (w/o Calib.).
Because simple spatial interpolation causes depth distortion, the reconstructed results do not align with the ground-truth reconstruction planes.
To identify the correct focal planes, the reconstructions for the calibrated bicubic baseline were evaluated at $4\times$ the original propagation distances.

Finally, we compared our network against H2HSR~\cite{no2024h2hsr}, a state-of-the-art deep learning framework originally designed for angle-of-view (AoV) expansion, which we adapted for volumetric spatial up-sampling.
We tested H2HSR using three distinct backbones: RDN (64 channels, 16 residual blocks), SwinIR, and HAT (180 dimensions).
To ensure a fair comparison, all DL baselines were trained using our patch-based cropping strategy rather than the full-resolution training proposed in the original H2HSR study.
Because the phase consistency loss used in H2HSR is incompatible with cropped patches, we replaced it with an $L_1$ fidelity loss while retaining their original perceptual loss.
Real-valued baseline models utilized the same initial learning rate ($4.0\times10^{-4}$) as our proposed network but required gradient normalization to maintain training stability.
To ensure stable convergence for transformer-based baselines (e.g., SwinIR and HAT), their initial learning rates were further lowered to $1.0\times10^{-4}$.

\paragraph{Optical configuration}

\begin{figure*}
    \centering
    \includegraphics[width=\linewidth]{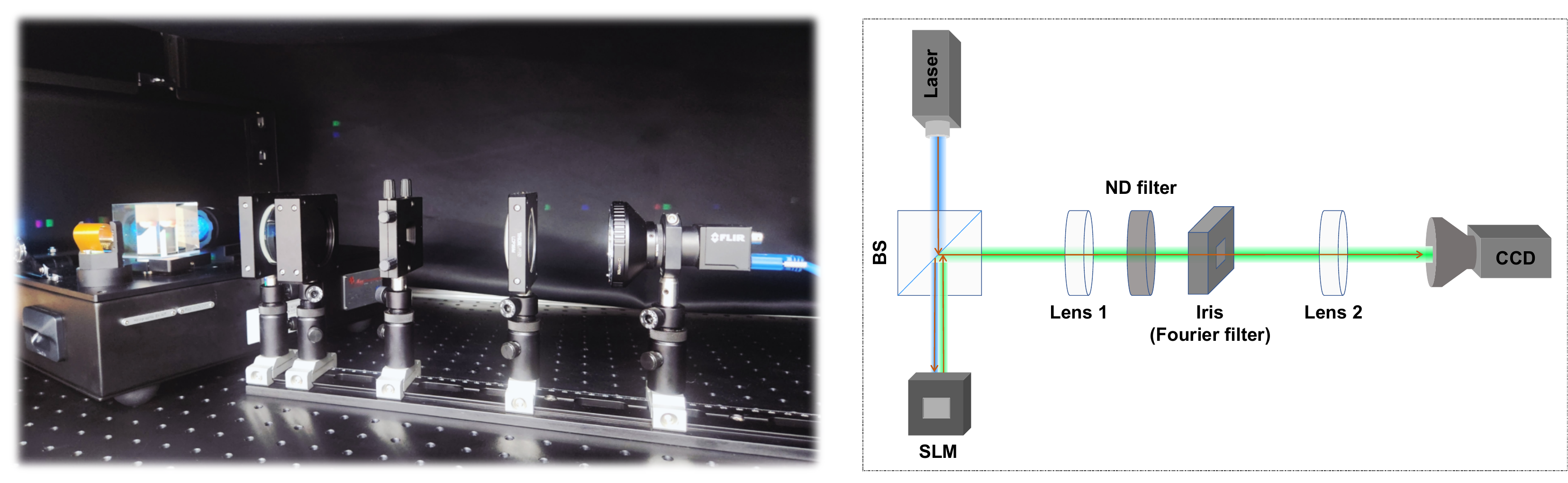}
    \caption{
    Optical system configuration for physical holographic reconstruction. The setup employs a $4f$ architecture with a focal length of 100~mm, utilizing an LCoS SLM illuminated by RGB lasers. An iris (OWIS SP60) is placed at the Fourier plane to filter specific diffraction orders, and the reconstructed scene is captured across multiple focal planes by a CCD camera (FLIR Blackfly S) mounted on a translation stage.
    }
    \label{fig:3_exp_optical_configuration}
\end{figure*}

For physical validation, we constructed a $4f$ optical system with a 100~mm focal length, as illustrated in Fig.~\ref{fig:3_exp_optical_configuration}.
The setup employs a holographic EV kit (May Inc.) featuring an LCoS SLM (IRIS-U62, $3840 \times 2160$ resolution) illuminated by RGB lasers (638~nm, 532~nm, and 450~nm).
Reconstructions were captured using a CCD camera (FLIR Blackfly S) mounted on a translation stage to evaluate multiple focal planes, with an iris placed in the Fourier plane to filter diffraction orders.

Complex-valued holograms were converted to phase-only representations using the double phase method (DPM)~\cite{maimone2017holographic}.
To match the physical near-plane of our system, the holograms were numerically propagated by an axial shift of $9.0$~mm prior to encoding. 
Furthermore, we applied a numerical off-axis carrier wave (shifted by 1.1 and 0.5 along the $x$- and $y$-axes, respectively) to suppress zero-order diffraction.
No iterative optimization was applied during encoding, ensuring the optical reconstructions strictly reflect the raw fidelity of the super-resolved holograms.
Finally, because a $1024 \times 1024$ resolution is physically too small for high-quality optical capture on our SLM, optical experiments were specifically conducted using $2048 \times 2048$ super-resolved holograms generated from $512 \times 512$ LR inputs.


\subsection{Evaluation}\label{subsec:evaluation}

\subsubsection{Quantitative evaluation}\label{subsec:quantitative_eval}
To quantitatively evaluate the super-resolved holograms, we assessed the reconstructed results using Peak Signal-to-Noise Ratio (PSNR), Structural Similarity Index (SSIM), and Learned Perceptual Image Patch Similarity (LPIPS)~\cite{ledig2017photo}.
For a comprehensive volumetric assessment, we numerically reconstructed each hologram at 40 uniformly spaced planes spanning from 0.0~mm to -7.3728~mm.
The metrics were calculated at each propagation plane and averaged to yield the final values presented in Table~\ref{tab:quantitative_all_models}.

\begin{table*}[t]
    \small
    \centering
    \begin{adjustbox}{width=\linewidth}
    \begin{tabular}{ccccccccccc}
        \toprule
        \multicolumn{2}{c}{\multirow{2}{*}{\textbf{Models}}} 
            & \multicolumn{3}{c}{\textbf{HologramSR}} 
            & \multicolumn{3}{c}{\textbf{BigBuckBunny}} 
            & \multicolumn{3}{c}{\textbf{RealSR}} \\
        \cmidrule(lr){3-5} \cmidrule(lr){6-8} \cmidrule(l){9-11}
        \multicolumn{2}{l}{} & PSNR(↑) & SSIM(↑) & LPIPS(↓) 
                             & PSNR(↑) & SSIM(↑) & LPIPS(↓)
                             & PSNR(↑) & SSIM(↑) & LPIPS(↓) \\
        \midrule
        \multirow{2}{*}{Bicubic} & \multicolumn{1}{|c|}{w/o Calib.}
            & 19.3689 & 0.3896 & 0.5760 
            & 23.6725 & 0.6775 & 0.4612 
            & 21.1183 & 0.4847 & 0.5455 \\
        & \multicolumn{1}{|c|}{w/ Calib.}
            & 19.4290 & 0.3423 & 0.5854 
            & 23.6153 & 0.5170 & 0.6025 
            & 21.2414 & 0.4446 & 0.5568 \\
        \midrule
        \multirow{3}{*}{H2HSR} & \multicolumn{1}{|c|}{RDN}
            & 28.0508 & 0.7347 & \secbest{0.3015}
            & 30.9585 & \secbest{0.8376} & 0.3148
            & \secbest{27.5971} & \secbest{0.7464} & 0.3331 \\
        & \multicolumn{1}{|c|}{SwinIR}
            & 27.7386 & 0.7168 & 0.3226
            & \secbest{30.9843} & 0.8347 & 0.3282
            & 27.5479 & 0.7359 & 0.3493 \\
        & \multicolumn{1}{|c|}{HAT}
            & \best{28.3336} & \secbest{0.7440} & \secbest{0.2926}
            & \best{30.9904} & \best{0.8380} & \secbest{0.3057}
            & \best{27.8073} & \best{0.7495} & \secbest{0.3283} \\
        \midrule
        \multirow{1}{*}{Ours} & \multicolumn{1}{|c|}{CV-RDN}
            & \secbest{28.1678} & \best{0.7497} & \best{0.2001}
            & 30.7333 & 0.8291 & \best{0.2318}
            & 27.3268 & 0.7324 & \best{0.3003} \\
        \bottomrule
    \end{tabular}
    \end{adjustbox}
    \caption{
    Quantitative comparison of volumetric super-resolution performance across three datasets.
    Bicubic interpolation is evaluated both with and without depth calibration, where calibration compensates for the quadratic depth distortion inherent to naive spatial up-sampling.
    H2HSR represents the baseline adapted with various deep-learning backbones (RDN, SwinIR, HAT), while our proposed network is denoted as CV-RDN.
    The best and second-best scores are marked in \textbf{bold} and \underline{underlined}, respectively.
    }
    \label{tab:quantitative_all_models}
\end{table*}

The baseline comparisons include bicubic interpolation and the H2HSR framework utilizing RDN, SwinIR, and HAT backbones.
As discussed in Sec.~\ref{sec:intro}, naive bicubic up-sampling of a hologram with a fixed pixel pitch introduces severe depth distortion, expanding the reconstruction volume quadratically.
To isolate this effect, we also report a calibrated version (\textbf{Bicubic*}), where the evaluation planes are scaled quadratically to align the focal depths (e.g., evaluating a 4.0~mm target plane at 16.0~mm for a $4\times$ spatial scale).
While calibration corrects the focal plane shift, the inherent depth-of-field (DoF) mismatches still degrade overall accuracy at unfocused planes (Fig.~\ref{fig:3_exp_qual_reconstruction_near_n_far_focus}).

As shown in Table~\ref{tab:quantitative_all_models}, the H2HSR-based models achieve highly competitive PSNR and SSIM scores, with the HAT variant yielding the highest overall fidelity metrics.
However, our proposed CV-RDN consistently achieves the best (lowest) LPIPS scores across all datasets.
This highlights a well-known trade-off in image restoration: while H2HSR's heavy reliance on an $L_1$ loss maximizes pixel-wise metrics (PSNR), our network's incorporation of the $\mathcal{L}_\mathrm{ASM-LPIPS}$ loss significantly enhances the perceptual and structural realism of the reconstructed 3D scenes.

\subsubsection{Qualitative evaluation}\label{subsec:qualitative_eval}

\begin{figure*}[t]
    \includegraphics[height=0.75\linewidth]{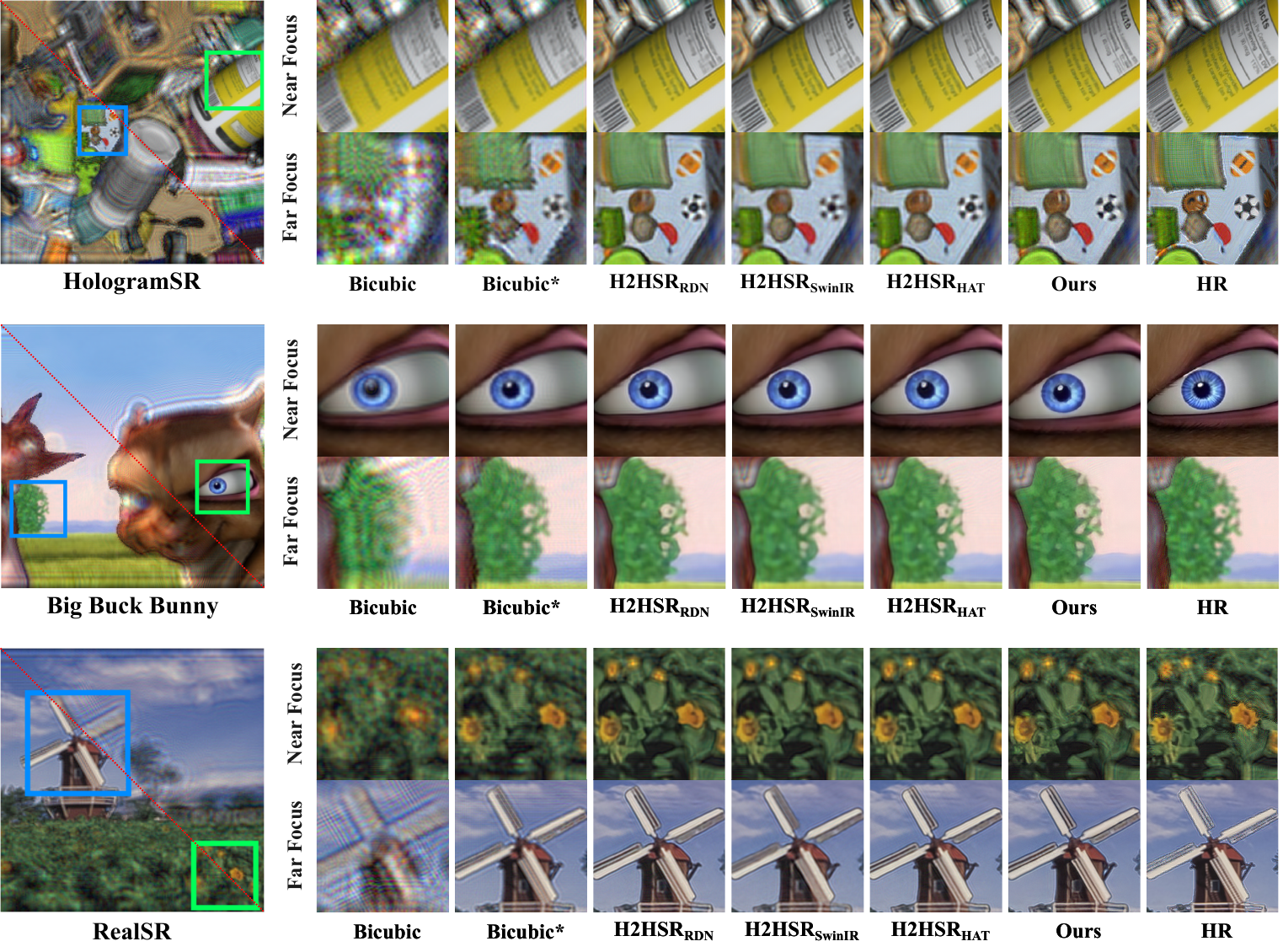}
    \caption{
    Qualitative comparison of reconstructed planes. The LR reference is divided by a red diagonal line into near-focus (upper, green box) and far-focus (lower, blue box) regions, corresponding to the top and bottom reconstruction rows, respectively.
    \textbf{Bicubic*} indicates calibrated up-sampling compensating for quadratic depth distortion. Our method achieves significantly sharper edges and textures than the alternatives.
    }
    \label{fig:3_exp_qual_reconstruction_near_n_far_focus}
\end{figure*}

The quantitative trade-off is visually evident in the reconstructed planes shown in Fig.~\ref{fig:3_exp_qual_reconstruction_near_n_far_focus}.
While naive bicubic interpolation completely fails to focus at the target distances due to depth distortion, the calibrated \textbf{Bicubic*} recovers the focus but lacks high-frequency details.
The H2HSR baselines successfully align the focal planes without calibration; however, they exhibit noticeable blurring artifacts characteristic of $L_1$-dominant optimization.
Interestingly, while H2HSR$_\mathrm{SwinIR}$ maintains high PSNR, it produces visually degraded, unnatural textures.
In contrast, our proposed model generates sharper contours and faithfully reproduces the high-frequency structural details present in the ground-truth HR holograms.

These qualitative results show consistent trends across datasets of varying complexity.
The Big Buck Bunny dataset presents relatively simple, synthetic scenes with lower structural complexity, whereas our HologramSR dataset features intricate characters and fine edges.
In the RealSR dataset, depth map quantization introduces imperfections in the HR reference, posing challenges for layer-based CGH generation.
Nevertheless, our proposed model demonstrates strong robustness against such reference degradation.
Notably, in the far-focus region (bottom row), our method reconstructs the windmill with sharper, more distinct structural details than both the H2HSR$_\mathrm{HAT}$ baseline and the imperfect ground truth.

\begin{figure*}[t!]
    \includegraphics[height=0.55\linewidth]{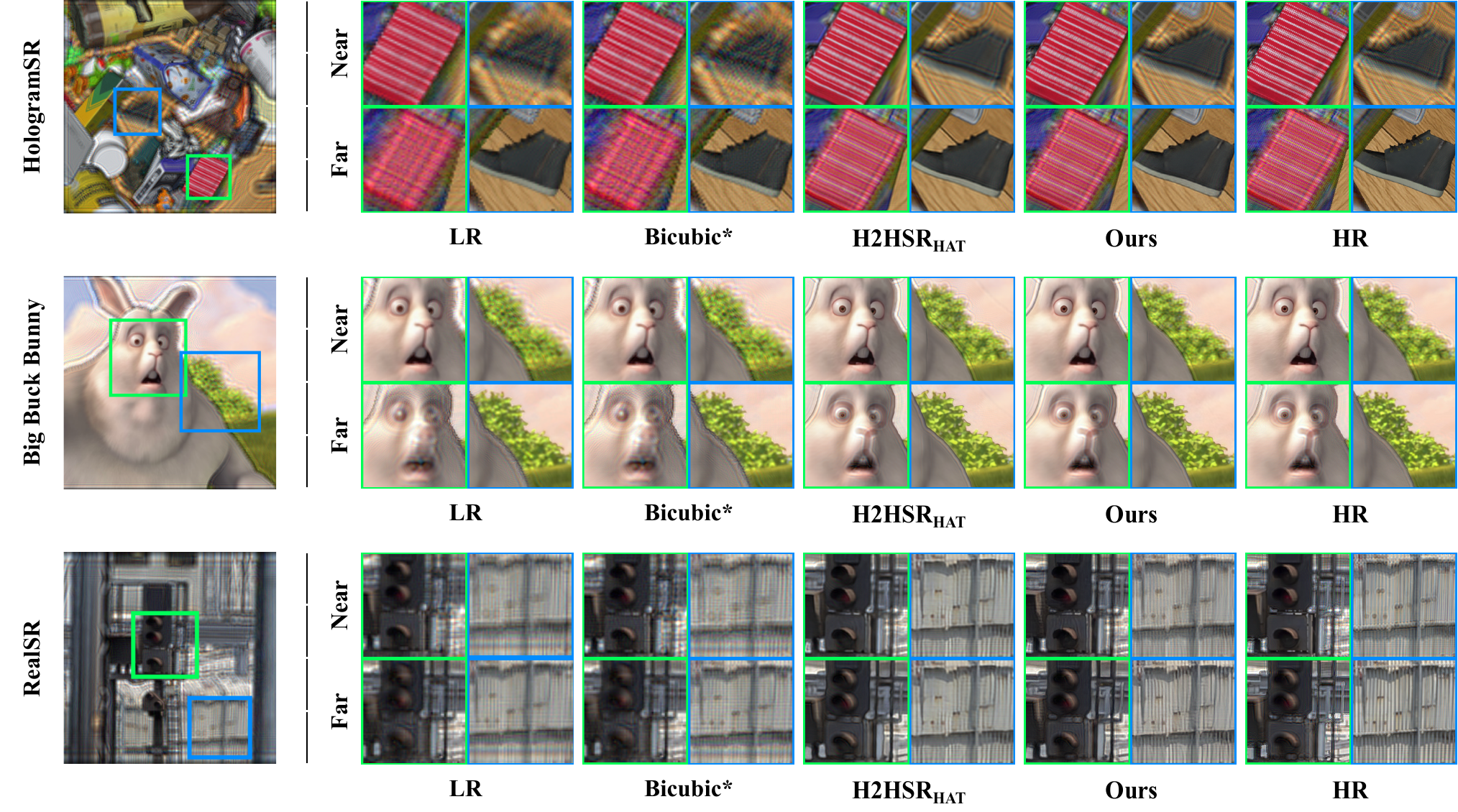}
    \caption{
    Visual quality comparison of focused and out-of-focus regions.
    The green and blue boxes highlight objects located at the near and far planes, respectively. 
    The \textbf{Near} and \textbf{Far} rows for each dataset represent numerical reconstructions at these respective focal distances for cross-validation.
    Our proposed method not only resolves sharp details for in-focus objects but also generates structurally consistent, natural defocus blur for out-of-focus fringe patterns.
    }
    \label{fig:3_exp_qual_reconstruction_separate}
\end{figure*}

This perceptual superiority extends to the representation of out-of-focus objects, as illustrated in Fig.~\ref{fig:3_exp_qual_reconstruction_separate}.
Accurate reproduction of defocus blur is essential for demonstrating the visual naturalness of 3D holographic displays.
In the LR reference, there is a clear distinction between focused and defocused regions. 
Calibrated bicubic interpolation appears visually similar to the LR input because it merely preserves the hologram's original, shallow depth-of-field (DoF) during up-sampling.
In contrast, deep learning approaches are required to achieve a linear depth scaling comparable to the HR ground truth.
Driven by our perceptual losses (Eq.~\ref{eq:asm-lpips-loss}), our model reconstructs high-quality details—such as the fine structural patterns of the tower in the HologramSR dataset—while precisely reproducing the fringe patterns originating from out-of-focus objects.
This structural stability is especially evident in the far-object regions of the RealSR dataset, proving that our approach provides clear perceptual advantages by generating visually natural and structurally consistent volumetric holograms.

\begin{figure*}[t!]
    \includegraphics[height=0.40\linewidth]{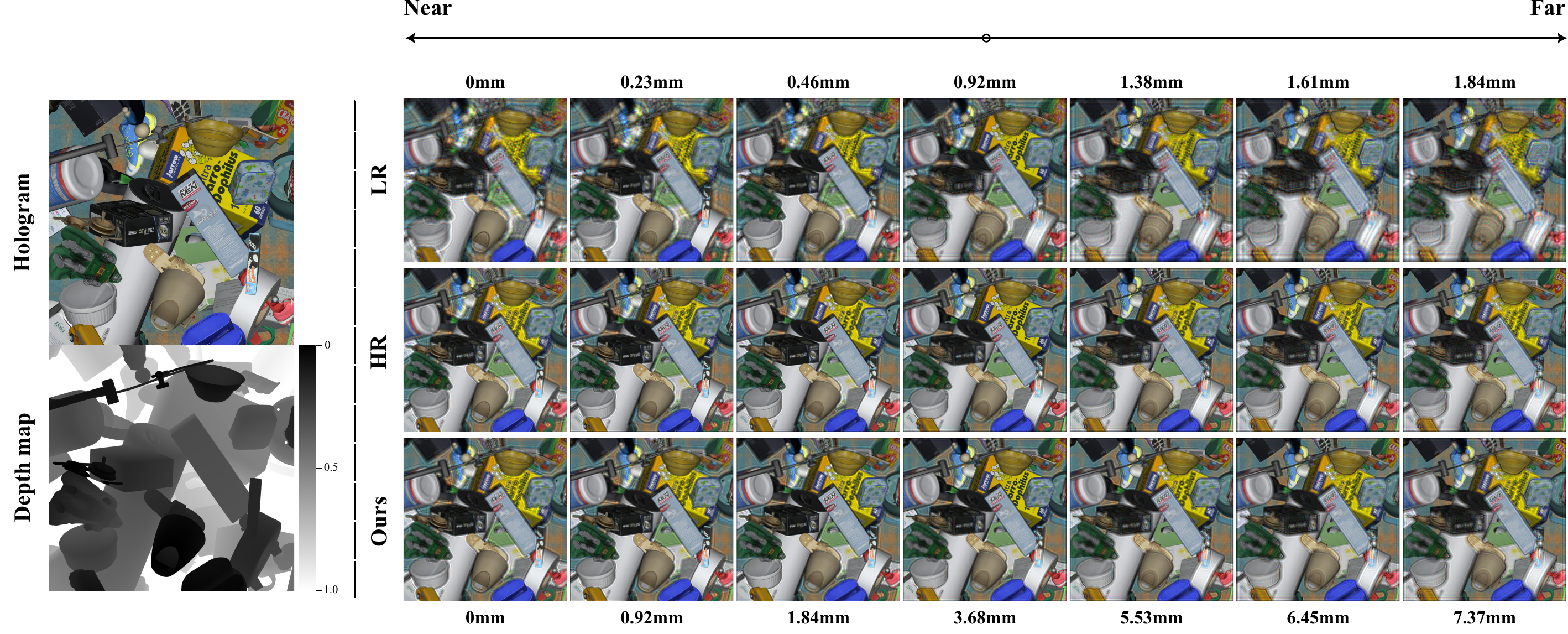}
    \caption{
    Continuous volumetric reconstruction sweep from the hologram plane (0.0~mm) to the maximum propagation distance.
    The RGB-D references are provided for color and depth cues.
    The proposed method successfully expands the shallow depth-of-field (DoF) of the LR input to match the linear DoF characteristics of the HR ground truth.
    }
    \label{fig:3_exp_qual_reconstruction_distances}
\end{figure*}

Finally, to validate that our model successfully learned the linear spatial-to-depth up-scaling relationship, we continuously sweep the reconstruction distance from 0.0~mm to the maximum depth.
As shown in Fig.~\ref{fig:3_exp_qual_reconstruction_distances}, the LR hologram exhibits a noticeably shallower DoF compared to the HR reference.
Our super-resolved hologram successfully expands this DoF, maintaining consistent focus alignment and blur intensity with the HR ground truth across the entire volumetric depth range.

\subsubsection{Effect of the Depth-Aware Perceptual Loss}

\begin{figure*}[t!]
    \includegraphics[height=0.43\linewidth]{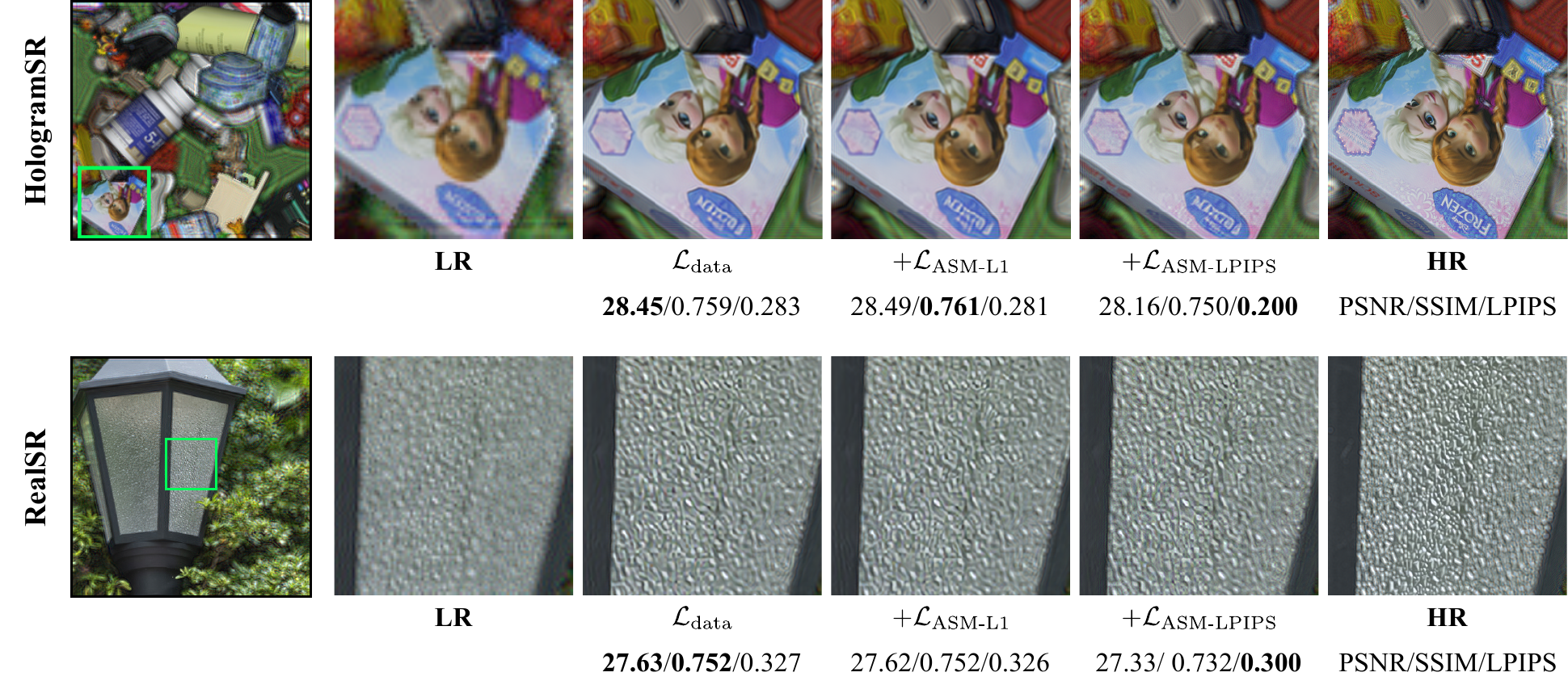}
    \caption{
    Qualitative and quantitative comparison of different loss function configurations on the HologramSR and RealSR datasets. 
    Enlarged regions are shown together with dataset-averaged PSNR/SSIM/LPIPS values for each method.
    }
    \label{fig:3_exp_ablation}
\end{figure*}

To isolate and verify the effect of our proposed objective function, we conducted an ablation study comparing different loss combinations. 
Fig.~\ref{fig:3_exp_ablation} presents qualitative reconstructions alongside dataset-averaged PSNR, SSIM, and LPIPS metrics for the HologramSR and RealSR datasets.
We evaluated the baseline data fidelity loss ($\mathcal{L}_{\mathrm{data}}$), the addition of an $L_1$-based reconstruction loss ($\mathcal{L}_{\mathrm{data}} + \mathcal{L}_{\mathrm{ASM\text{-}L1}}$), and our proposed full configuration ($\mathcal{L}_{\mathrm{data}} + \mathcal{L}_{\mathrm{ASM\text{-}LPIPS}}$).

As demonstrated in Fig.~\ref{fig:3_exp_ablation}, adding the $\mathcal{L}_{\mathrm{ASM\text{-}L1}}$ term to the baseline provides only marginal changes to the metrics and visual quality.
This indicates that applying a simple pixel-wise $L_1$ penalty in the reconstructed wavefield domain has a limited effect on resolving fine details. 

In contrast, replacing it with the depth-aware perceptual loss ($\mathcal{L}_{\mathrm{ASM\text{-}LPIPS}}$) triggers a distinct perception-distortion trade-off.
While PSNR and SSIM decrease slightly compared to the purely $L_1$-driven approaches, the LPIPS score improves significantly.
Qualitatively, this translates to visually sharper results; fine patterns and structural contours are distinctly restored in both datasets.
This confirms that evaluating perceptual features (via LPIPS) is vastly more effective than pixel-wise regression ($L_1$) for preserving high-frequency interference patterns in holography.
Furthermore, this finding perfectly explains the behavior of the H2HSR baseline evaluated earlier: because H2HSR relies heavily on an $L_1$-based reconstruction loss, it inherently favors over-smoothed solutions, whereas our method successfully maintains perceptual realism.

\subsubsection{Optical reconstruction evaluation}
\begin{figure*}[t!]
    \includegraphics[height=0.85\linewidth]{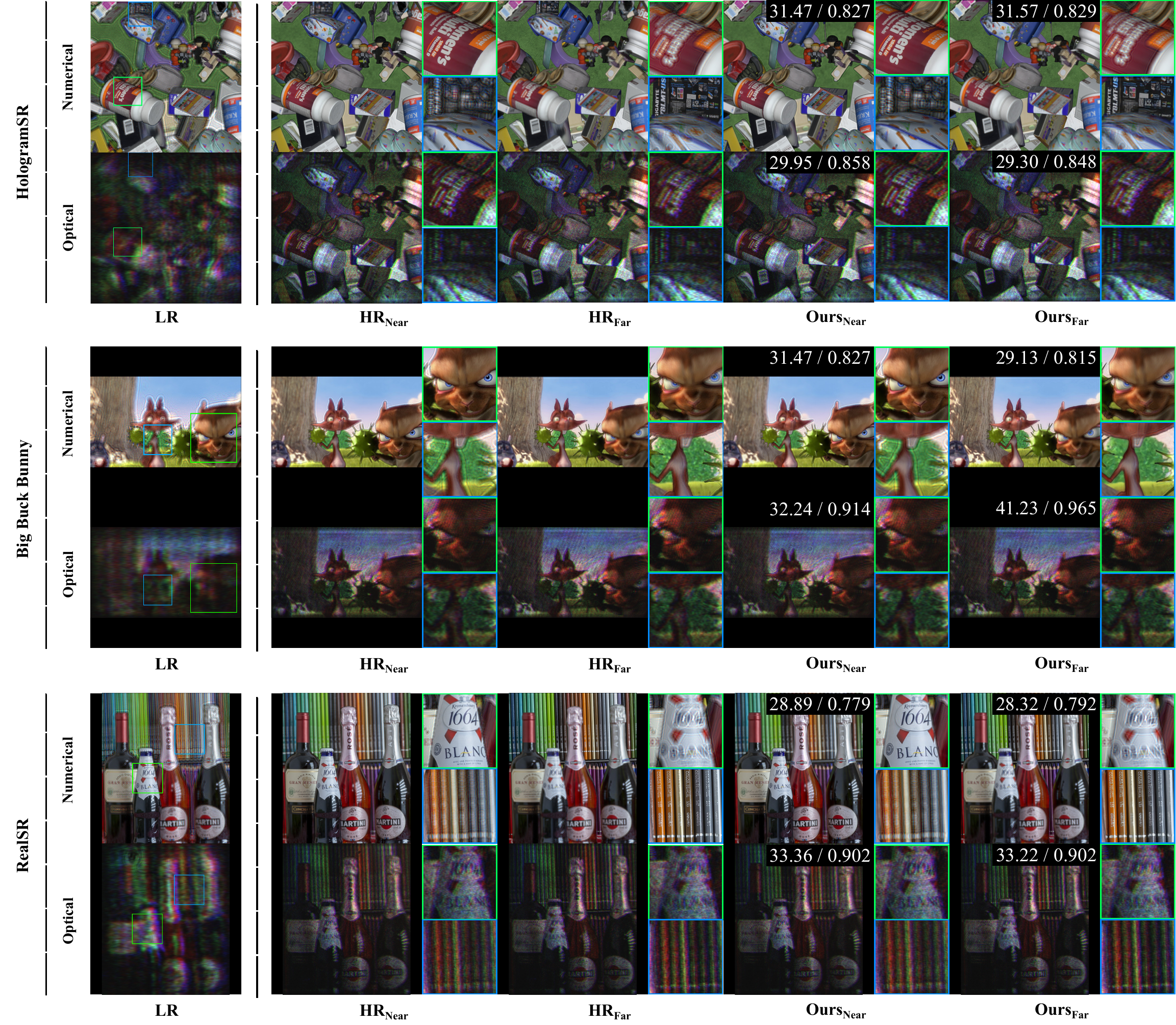}
    \caption{
    Optical and numerical reconstruction results across the HologramSR, Big Buck Bunny, and RealSR datasets.
    The green and blue boxes highlight magnified regions corresponding to objects focused at the near and far planes, respectively.
    For our proposed method, quantitative metrics (PSNR and SSIM, computed in the captured image domain) are displayed directly on the results.
    }
    \label{fig:3_exp_optical_reconstruction}
\end{figure*}

Fig.~\ref{fig:3_exp_optical_reconstruction} presents the physical optical reconstruction results.
To validate volumetric accuracy, we compared both numerical and optical reconstructions at the nearest and farthest focal planes.
While the LR holograms exhibit recognizable structures during numerical simulation, their physical optical reconstructions suffer from severe degradation, making it impossible to isolate clear focal planes.
In contrast, our super-resolved holograms successfully project distinct, high-contrast images that closely mirror the HR ground truth.
Notably, the quantitative PSNR and SSIM values of our optical reconstructions are highly comparable to those of the actual HR holograms.

Translating complex-valued holograms to physical displays inevitably introduces hardware-specific artifacts.
Specifically, phase-only quantization from DPM encoding and off-axis spatial filtering cause minor brightness variations and background noise.
While near-plane objects are restored with high fidelity, far-plane reconstructions exhibit slightly more degradation due to longer propagation distances compounding these optical constraints.
However, because these physical limitations equally affect the HR ground-truth holograms, the relative optical performance of our proposed method remains robust.

Overall, these physical experiments confirm the practical viability of our network.
Even under strict hardware constraints and encoding losses, the proposed method generates super-resolved holograms with an optical perceptual quality that is nearly indistinguishable from the target ground truth, proving that our up-sampling approach successfully maps to real-world wave propagation.


\subsection{Adaptation to Unseen Depth Ranges}\label{subsec:result_LORA}

\begin{figure*}[t]
    \centering
    \includegraphics[width=\linewidth]{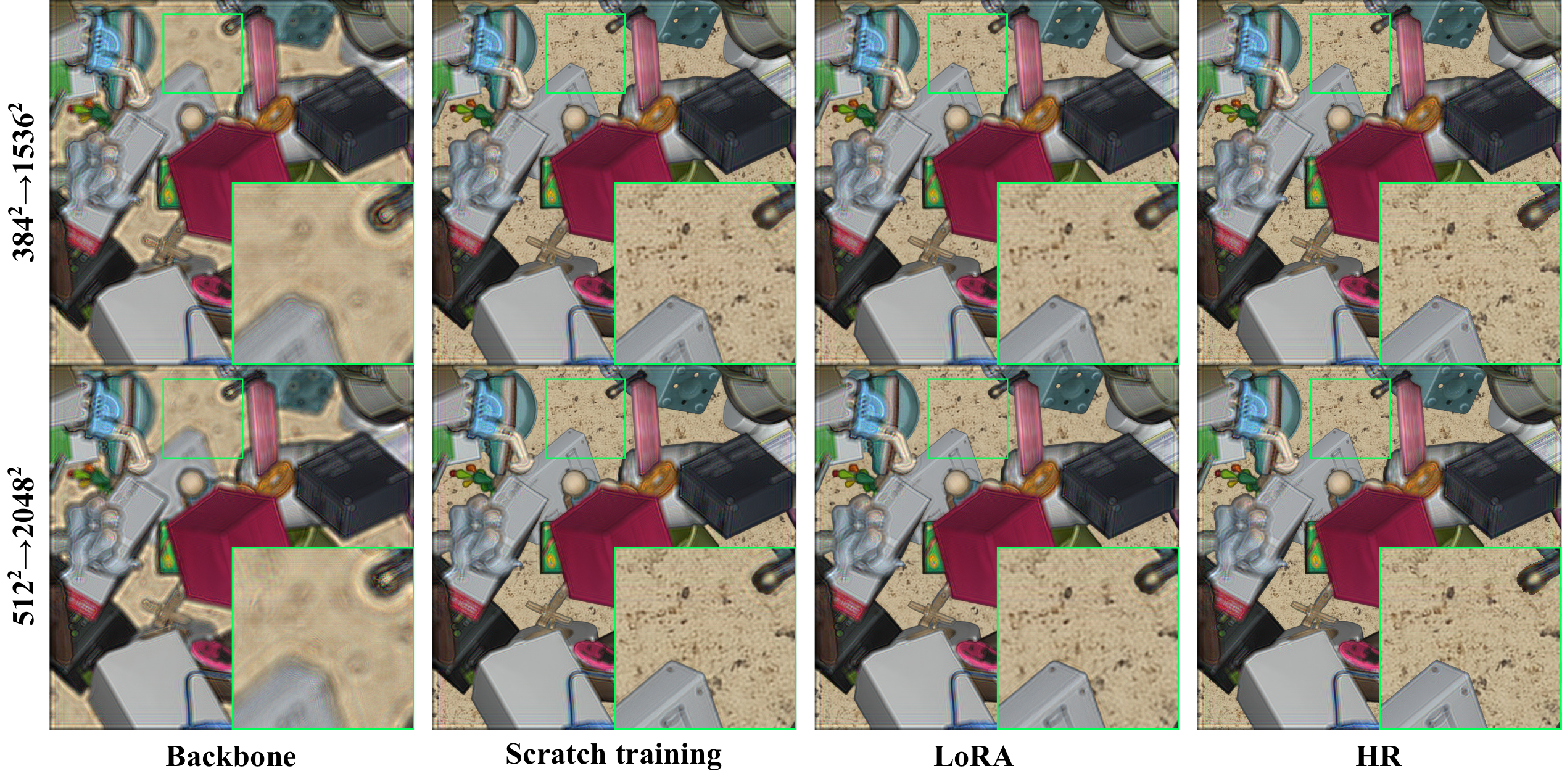}

    \vspace{0.8em}

    \small
    \begin{adjustbox}{width=\linewidth}
    \begin{tabular}{c cccc ccc}
        \toprule
        \multirow{2}{*}{\textbf{Methods}}
            & \multicolumn{3}{c}{\textbf{384$\rightarrow$1536}} 
            & \multicolumn{3}{c}{\textbf{512$\rightarrow$2048}}
            & \multirow{2}{*}{\textbf{Train Time (h)}} \\
        \cmidrule(lr){2-4} \cmidrule(lr){5-7}
            & PSNR($\uparrow$) & SSIM($\uparrow$) & LPIPS($\downarrow$)
            & PSNR($\uparrow$) & SSIM($\uparrow$) & LPIPS($\downarrow$) & \\
        \midrule
        Backbone
            & 25.2084 & 0.6776 & 0.2876
            & 24.0717 & 0.6538 & 0.3409
            & 22.5 \\
        Scratch training
            & \secbest{29.1112} & \secbest{0.7629} & \secbest{0.2211}
            & \secbest{29.3594} & \secbest{0.7662} & \secbest{0.2472}
            & 22.5 \\
        LoRA$_\mathrm{D50}$
            & 28.8906 & 0.7562 & 0.2286
            & 28.8948 & 0.7536 & 0.2621
            & 1.2 \\
        LoRA$_\mathrm{D100}$
            & 29.0594 & 0.7618 & 0.2225
            & 29.1925 & 0.7624 & 0.2520
            & 2.5 \\
        LoRA$_\mathrm{D200}$
            & \best{29.2019} & \best{0.7668} & \best{0.2170}
            & \best{29.4292} & \best{0.7700} & \best{0.2428}
            & 5.2 \\
        \bottomrule
    \end{tabular}
    \end{adjustbox}

    \caption{Comparisons to evaluate depth-range adaptation under LoRA-based fine-tuning. The upper panel shows qualitative results, and the lower table reports quantitative comparisons between standard training and LoRA-based training under two super-resolution settings. The best and second-best results are highlighted in bold and underline, respectively.
    }
    \label{fig:3_exp_lora}
\end{figure*}

To evaluate the parameter-efficient adaptation strategy proposed in Sec.~\ref{sec:LoRA}, we conducted experiments on higher-resolution holograms that inherently span extended depth ranges, introducing new fringe statistics.
Two up-sampling scenarios were tested: $384^2 \rightarrow 1536^2$ and $512^2 \rightarrow 2048^2$.
Following our methodology, LoRA modules (with rank $r=8$ and scaling factor $\alpha=16$) were injected into the complex-valued convolutions of the backbone, which was pre-trained on the $256^2 \rightarrow 1024^2$ task.
To assess data and computational efficiency, fine-tuning was performed using extremely limited subsets of 50, 100, and 200 samples (denoted as $\mathrm{LoRA}_{D50}$, $\mathrm{LoRA}_{D100}$, and $\mathrm{LoRA}_{D200}$), alongside a baseline network fully trained from scratch.

Because these higher-resolution tasks demand reconstruction across extended depth ranges, direct inference using the unmodified backbone suffers heavily from the encoder depth bias identified earlier.
This out-of-distribution degradation is most pronounced in the $512^2 \rightarrow 2048^2$ scenario, where the target volume extends furthest beyond the pre-training limits.
However, by fine-tuning only our strategically placed complex-valued low-rank parameters, the network successfully re-calibrates its depth-dependent mapping.
As demonstrated by the qualitative results in Fig.~\ref{fig:3_exp_lora}, the adapted model accurately resolves the maximum-depth regions, fully overcoming the inherent depth bias to match the perceptual fidelity of the scratch-trained model.

Furthermore, despite utilizing only a fraction of the training data, the quantitative results confirm that the adapted models achieve performance highly comparable to networks trained entirely from scratch.
Notably, $\mathrm{LoRA}_{D200}$ even slightly outperforms the full scratch-training baseline in both resolution settings.
This data efficiency translates directly to massive computational savings: while training from scratch requires approximately 22.5 hours, adapting the model via $\mathrm{LoRA}_{D200}$ requires only 5.2 hours. 
Overall, these results prove that our tailored, complex-valued LoRA design is a highly effective, time-saving solution for adapting a single pre-trained super-resolution backbone to novel optical configurations and unseen depth ranges.


\section{Conclusion}

In this study, we proposed a novel complex-valued hologram super-resolution framework designed specifically for the volumetric up-sampling of three-dimensional scenes.
Unlike previous methods, our approach successfully preserves the linear depth scaling inherently required for physically consistent 3D holographic displays.
By leveraging a Complex-Valued Residual Dense Network (CV-RDN) optimized with a depth-aware perceptual reconstruction loss, our model effectively overcomes the over-smoothing and high-frequency degradation typical of conventional complex-domain pixel-wise regression. 

Furthermore, we addressed the inherent depth bias of pre-trained convolutional encoders by introducing a parameter-efficient fine-tuning strategy.
By strategically injecting complex-valued Low-Rank Adaptation (LoRA) modules into the backbone network, we demonstrated a highly data- and time-efficient method for adapting the model to unseen depth ranges and higher spatial resolutions.
Both comprehensive numerical simulations and physical optical experiments confirmed that our proposed framework delivers superior perceptual fidelity and structural consistency across expansive depth volumes.

\paragraph{Limitation and Future Work}
While complex-valued convolution layers provide a physically principled and highly effective representation for holographic data, their computational cost remains a notable limitation.
Even though our complex-valued architecture halves the channel dimensions to match the parameter count of real-valued models, the underlying decomposition into multiple real-valued operations results in slower inference speeds.
To facilitate the real-time deployment of deep neural networks in practical holographic displays, future work will focus on improving computational efficiency through network quantization and the development of more streamlined complex-valued operators.

Additionally, while our LoRA-based fine-tuning provides an effective, parameter-efficient solution for adapting to new physical configurations, achieving fundamental zero-shot depth generalization remains an open challenge.
Ultimately, our goal is to develop architectures capable of intrinsically learning the generalized physical relationship between hologram fringe statistics and infinite depth ranges. This would enable robust, artifact-free inference on out-of-range holograms without the need for supplementary adaptation datasets or fine-tuning stages.

\bibliographystyle{elsarticle-num}
\bibliography{reference}
\end{document}